\documentclass[12pt,fleqn]{article}
\usepackage{amssymb}
\usepackage{hyperref}
\usepackage{amsmath}

\usepackage{amsmath,amssymb,amsthm}

\begin{document}
\begin{center}
{\Large \textbf{Eikonal 3.1: General solution for coupled system of eikonal equations, three space dimensions}}

\vskip 20pt {\large \textbf{Iryna YEHORCHENKO}}

\vskip 20pt {Institute of Mathematics of NAS Ukraine, 3 Tereshchenkivs'ka Str., Kyiv-4, Ukraine} \\
E-mail: iyegorch@imath.kiev.ua
\\
Institute of Mathematics, PAN, Poland\\
iyehorchenko@impan.pl
\end{center}

\footnotesize
\begin{abstract}
A general solution for a coupled system of eikonal equations
$$
u_\mu u_\mu =0,
$$
$$
v_\mu v_\mu =0,
$$
$$
u_\mu v_\mu =1
$$
is presented, where lower indices designate derivatives, $\mu=0,1,2,3$, and summation is implied over the repeated indices.

This solution is of interest by itself due to wide applications of the eikonal equations, but the system considered also appears to be part of the reduction conditions for many equations of mathematical physics.

\end{abstract}

\normalsize
\section{Background Ideas}
Eikonal equations and their systems are used in problems of wave propagation, physical (wave) and geometric (ray) optics. For more information on the eikonal equation and its general solutions and symmetry
see \cite{preprintIY2022Eik}.

However, I became interested in these equations as they appear in reduction conditions of partial differential equations.

Reduction of a certain PDE e.g. of the form
$$\Phi(x,u,u_x,u_{xx})$$ (here we designate by $x$, $u$ sets
 of independent and dependent variables, and by $u_x$ etc. sets of the respective partial derivatives) is substitution of certain ansatz
e.g. of the form
$$u=\phi(\omega_1,\omega_2,...,\omega_k),$$
where $\phi$ are new dependent variables, and $\omega_i$ are
new independent variables, with the number of these
new variables smaller than the number of old variables.

The ansatz will produce a reduction if the substitution of this ansatz into the original equation will give a new (ODE or PDE) equation (or a new system of equations) with smaller number of independent variables.

Ansatzes cannot be arbitrary. The Lie symmetry analysis with classification of inequivalent subalgebras of the Lie symmetry algebra produces Lie symmetry ansatzes. Lie symmetry gives not all possible ansatzes and whence
not all possible reductions. Direct method involves direct substitution of an arbitrary ansatz
of a certain form into an equation and finding and solution of reduction conditions.
not all possible reductions. Eikonal equations, and in some cases (when we consider reduction to two new variables) coupled eikonal equations,
appear as reduction conditions for all Poincar\'e-invariant equations.

Here we will consider the system
\begin{gather}
u_\mu u_\mu =0,\\ \nonumber
v_\mu v_\mu =0,\\ \label{eik1}
u_\mu v_\mu =1, \nonumber
\end{gather}
\noindent
where lower indices designate derivatives, $\mu=0,1,2,3$ and summation in Minkowski space is implied over the repeated indices.

In \cite{preprintIY2017} we found a rank 1 parametric solution for $u$ and $v$ and two space variables:
\begin{gather} \nonumber
u=\frac{x_1+\frac{x_2 z}{\sqrt{1-z^2}}-k'(z)}{g'(z)},
\end{gather}
\begin{gather} \nonumber
v=\frac{gx_2}{\sqrt{1-z^2}}+
\frac{p(z)}{g'(z)}[x_1+\frac{x_2z}{\sqrt{1-z^2}}-k'(z)]+r(z),
\end{gather}
\begin{gather} \nonumber
0=x_0-x_1z+x_2\sqrt{1-z^2}+
\frac{g(z)}{g'(z)}\{x_1+\frac{x_2 z}{\sqrt{1-z^2}}-k'(z)\}-h(z)
\end{gather}
Here
\begin{gather} \nonumber
r'=-k''\times(zg+(1-z^2)g'),\\ \nonumber
p=\frac{1}{2}\{-g'^2+(g-zg')^2\}.
\end{gather}

\section{Use of Hodograph and Contact Transformations}
Here we consider in detail the process for finding of the rank 2 parametric solution, depending on two parameter. Our initial dependent functions are
$$u=u(x_0,x_1,x_2,x_3), \ v=v(x_0,x_1,x_2,x_3).$$
The method we use here was developed on the basis of ideas presented in the paper by Zhdanov, Revenko and Fushchych \cite{FZhR-preprint} on the general solution of the d'Alembert-Hamilton system, however, we use somewhat different formulas.
We go to new dependent variables $w$ and $v$, and new independent variables $y_0$, $y_1$, $y_2$, $y_3$.

\begin{equation} \nonumber
u=y_0,\quad x_0=v, \quad x_1=y_1, \quad x_2=y_2, \quad x_3=y_3.
\end{equation}

Expressions for derivatives
\begin{gather} \nonumber
u_{x_0}=\frac{1}{w_{y_0}},\qquad
u_{x_1}=-\frac{w_{y_1}}{w_{y_0}},\qquad
u_{x_2}=-\frac{w_{y_2}}{w_{y_0}},\qquad
u_{x_3}=-\frac{w_{y_3}}{w_{y_0}},\\ \nonumber
v_{x_0}=\frac{v_{y_0}}{w_{y_0}},\qquad
v_{x_1}=v_{y_1}-v_{y_0}\frac{w_{y_1}}{w_{y_0}},\\ \nonumber
v_{x_2}=v_{y_2}-v_{y_0}\frac{w_{y_2}}{w_{y_0}},\qquad
v_{x_3}=v_{y_3}-v_{y_0}\frac{w_{y_3}}{w_{y_0}}. \nonumber
\end{gather}

Please note that in the new equations obtained after the hodograph transformation we denote $v_{y_\mu}$
as $v_\mu$, and $w_{y_\mu}$ as $w_\mu$.

Substitution to the first equation of (\ref{eik1}) gives
\begin{gather}
-\frac{w_1^2}{w_0^2}-\frac{w_2^2}{w_0^2}-\frac{w_3^2}{w_0^2}+\frac{1}{w_0^2}=0 \label{eik2.1}
\end{gather}
As we assumed $w_0\neq 0$, this equation is equivalent
to
\begin{equation}\nonumber
w_1^2+w_2^2+w_3^2=1
\end{equation}

Substitution to the second equation of (\ref{eik1})
\begin{gather}\nonumber
v_1^2+v_0^2 \frac{w_1^2}{w_0^2}
-2\frac{v_0 v_1 w_1}{w_0}+ v_2^2+
v_0^2 \frac{w_2^2}{w_0^2}-2\frac{v_0v_2w_2}{w_0}
+ v_3^2+v_0^2 \frac{w_3^2}{w_0^2} -2\frac{v_0v_3w_3}{w_0}= \frac{v_0^2}{w_0^2}
\end{gather}

\begin{gather}\nonumber
v_1^2+ v_2^2+ v_3^2-2(v_1 w_1+v_2 w_2+v_3 w_3)\frac{v_0}{w_0}= 0
\end{gather}

Substitution to the third equation of (\ref{eik1})
\begin{gather}\nonumber
\frac{v_0}{w_0^2}+\frac{w_1}{w_0}(v_1+v_0\frac{w_1}{w_0})
+\frac{w_2}{w_0}(v_2+v_0\frac{w_2}{w_0})
+\frac{w_3}{w_0}(v_3+v_0\frac{w_3}{w_0})=1
\end{gather}

As $w_0\neq 0$, we can have
\begin{gather} \nonumber
v_1 w_1+v_2 w_2+v_3 w_3=w_0
\end{gather}

Then the third equation of (\ref{eik1}) takes the form
\begin{equation} \nonumber
v_1^2+v_2^2+v_3^2=v_0
\end{equation}

The resulting system
\begin{gather}\label{eik4}
w_{y_1}^2+w_{y_2}^2+w_{y_3}^2=1 \\ \label{eik4a}
v_{y_1}^2+v_{y_2}^2+v_{y_3}^2=v_0 \\ \label{eik4b}
v_{y_1} w_{y_1} +v_{y_2} w_{y_2}+v_{y_3} w_{y_3}=w_{y_0}.
\end{gather}

\section{Contact Transformations}
New independent variables
$z_0=y_0$, $z_1=w_{y_1}$,  $z_2=w_{y_2}$, $z_3=y_3$

New dependent variable $H=y_1w_{y_1}+y_2w_{y_2}-w$

Relations for derivatives with respect to new independent
variables
\begin{gather} \nonumber
H_{z_0}=-w_{y_0}, \quad H_{z_1}=y_1, \quad H_{z_2}=y_2, \quad H_{z_3}=-w_{y_3},\\ \nonumber
v_{y_0}=v_{z_0}+v_{z_1}w_{y_0 y_1}+v_{z_2}w_{y_0 y_2},\quad
v_{y_1}=v_{z_1}w_{y_1 y_1}+v_{z_2}w_{y_1 y_2},\\ \nonumber
v_{y_2}=v_{z_1}w_{y_1 y_2}+v_{z_2}w_{y_2 y_2},\quad
v_{y_3}=v_{z_3}+v_{z_1}w_{y_1 y_3}+v_{z_2}w_{y_2 y_3}.
\end{gather}
\begin{gather} \nonumber
w_{y_1 y_1}=\frac{H_{z_2 z_2}}{D_{12}},\quad
w_{y_1 y_2}=-\frac{H_{z_1 z_2}}{D_{12}},\quad
w_{y_2 y_2}=\frac{H_{z_1 z_1}}{D_{12}},\\ \nonumber
D_{ab}=\left|
  \begin{array}{cc}
    H_{z_a z_a} & H_{z_a z_b} \\
    H_{z_a z_b} & H_{z_b z_b} \\
  \end{array}
\right|,
\end{gather}
\begin{gather} \nonumber
w_{y_0 y_1}=-\frac{D_{02}}{D_{12}},\quad
w_{y_0 y_2}=\frac{D_{01}}{D_{12}},\quad
w_{y_1 y_3}=\frac{D_{13}}{D_{12}},\quad
w_{y_2 y_3}=-\frac{D_{23}}{D_{12}},\\ \nonumber
D_{0a}=\left|
  \begin{array}{cc}
    H_{z_0 z_1} & H_{z_0 z_2} \\
    H_{z_1 z_a} & H_{z_2 z_a} \\
  \end{array}
\right|. \nonumber
\end{gather}

\section{Substitution of Contact Transformations into System (\ref{eik4}-\ref{eik4b})}
The first equation results in
\begin{equation} \nonumber
z_1^2+z_2^2+H_{z_3}^2=1
\end{equation}
\noindent
that has a general solution for the function $H$
\begin{equation}\label{H}
H=z_3\sqrt{1-z_1^2-z_2^2}+G(z_0,z_1,z_2),
\end{equation}
where $G$ is an arbitrary function of its arguments.

Substitution of the contact transformations into  (\ref{eik4b}) gives
\begin{gather}\nonumber
(v_{z_1}w_{y_1 y_1}+v_{z_2}w_{y_1 y_2})w_{y_1}+
(v_{z_1}w_{y_1 y_2}+v_{z_2}w_{y_2 y_2})w_{y_2}+
(v_{z_3}+v_{z_1}w_{y_1 y_3}+
v_{z_2}w_{y_2 y_3})w_{y_3}= \\ \nonumber
v_{z_1}z_1\frac{H_{z_2 z_2}}{D_{12}}-
v_{z_2}z_1\frac{H_{z_1 z_2}}{D_{12}}-
v_{z_1}z_2\frac{H_{z_1 z_2}}{D_{12}}+
v_{z_2}z_2\frac{H_{z_1 z_1}}{D_{12}} + \\ \nonumber
(v_{z_3}+v_{z_1}\frac{D_{13}}{D_{12}}+
v_{z_2}\frac{D_{23}}{D_{12}})H_{z_3}
=-H_{z_0}
\end{gather}

From (\ref{H})
\begin{gather} \nonumber
H_{z_0}=G_{z_0}, \quad H_{z_3}=\sqrt{1-z_1^2-z_2^2}=s, \quad
H_{z_1}=-\frac{z_1 z_3}{s}+G_{z_1}, \quad
H_{z_2}=-\frac{z_2 z_3}{s}+G_{z_2}, \\ \nonumber
H_{z_0 z_1}=G_{z_0z_1}, \quad H_{z_0 z_2}=G_{z_0z_2}, \quad
H_{z_1 z_2}= -\frac{z_1z_2z_3}{s^{\frac{3}{2}}}+G_{z_1z_2}, \\ \nonumber
H_{z_1 z_1}=-\frac{z_3}{s}-
\frac{z_1^2z_3}{s^{\frac{3}{2}}} + G_{z_1 z_1},\quad
H_{z_2 z_2}=-\frac{z_3}{s}-
\frac{z_2^2z_3}{s^{\frac{3}{2}}} + G_{z_2 z_2}.
\end{gather}

Then substitution of the contact transformations into (\ref{eik4b}) gives
\begin{equation} \nonumber
G_{z_0}-v_{z_3}s=0
\end{equation}
that gives an expression for $v$
\begin{equation}\label{v1}
v=\frac{G_{z_0}z_3}{s}+P(z_0,z_1,z_2),
\end{equation}
where $P(z_0,z_1,z_2)$ is an arbitrary function of its arguments.

From (\ref{v1}) and (\ref{eik4a}), (\ref{eik4b})  we get
\begin{gather}\nonumber
G_{z_0}=g(z_1,z_2), \quad  P_{z_0}=p(z_1,z_2)
\end{gather}

Whence we get we get a parametric general solution for the system (\ref{eik4})-(\ref{eik4b})
\begin{gather*}
v=\frac{gy_3}{s}+py_0+r, \\
w=y_1 z_1+y_2 z_2 -y_3 s-gy_0 -k, \label{vw} \\
0=y_1+\frac{y_3z_1}{s}-g_{z_1}y_0 -k_{z_1}, \\ \nonumber
0=y_2+\frac{y_3z_2}{s}-g_{z_2}y_0 -k_{z_2}. \nonumber
\end{gather*}

Applying transformations inverse to the hodograph transformations, we get a parametric general solution of the coupled eikonal system (\ref{eik1}) for the original functions $u$ and $v$:
\begin{gather*}
u=\frac{1}{g}(x_1z_1+x_2z_2 -x_3s -x_0+k),
\\
v=\frac{gx_3}{s}+ pu+r,
\\
0=x_1-\frac{x_3z_1}{s}-g_{z_1}u -k_{z_1}
\\
0=x_2-\frac{x_3z_2}{s}-g_{z_2}u -k_{z_2}.
\end{gather*}
\noindent
where
\begin{gather*}
r_{z_1}=-k_{z_1z_1}(z_1g+s^2g_{z_1}),\\
r_{z_2}=-k_{z_1z_2}(z_2g+s^2g_{z_2}),\\
p=\frac{1}{2}(-(g_{z_1}^2+g_{z_2}^2-g_{z_1}g_{z_2})+
(g-z_1g_{z_1}-z_2g_{z_2})^2),
\end{gather*}
\noindent
$g$ and $k$ are arbitrary functions of its arguments
$z_1$ and $z_2$. Note that we skipped consideration of solutions with ranks 0 and 1, that can be written similarly to solutions for the system with two space variables found in \cite{preprintIY2017}.

\section{Conclusions}
The general parametric solutions presented can be used for solution of boundary and initial problems, for testing of numerical methods and to find new exact solutions of Poincar\'e invariant equations, that cannot be found by the standard symmetry methods.

\section{Acknowledgements}
The first and foremost my acknowledgements go to the
Armed Forces of Ukraine and to the Territorial Defence Forces of Ukrainian Regions due to whom I am alive and is able to work.

Please remember that Russia is an aggressor country and still plans to kill all Ukrainians not going to be their subordinates, and Russian scientists contribute to killings despite words about peace from a tiny portion of them; if even by trying to persuade na\"{i}ve people that Russia is a civilised country.

This research was mostly completed during my work at the Institute of Mathematics of the Polish Academy of Sciences with the grant of the Narodowe Centrum Nauki (Poland), Grant No.2017/26/A/ST1/00189. I would like to thank the Institute of Mathematics of the Polish Academy of Sciences for their hospitality and grant support, to the National Academy of Sciences of the USA and the National Centre of Science of Poland for their grant support.

Further work was supported by a grant from the Simons Foundation (1290607, I.A.Y).

\end{document}